\documentclass[%
aps,
rsi,
amsmath,
amssymb,
superscriptaddress,
reprint,%
]{revtex4-2}

\usepackage{graphicx}
\usepackage{dcolumn}
\usepackage{bm}

\usepackage[utf8]{inputenc}
\usepackage[T1]{fontenc}
\usepackage{mathptmx}
\usepackage{etoolbox}
\usepackage[color=lightgray]{todonotes}
\usepackage{siunitx}
\usepackage{amsmath}

\makeatletter
\def\@email#1#2{%
 \endgroup
 \patchcmd{\titleblock@produce}
  {\frontmatter@RRAPformat}
  {\frontmatter@RRAPformat{\produce@RRAP{*#1\href{mailto:#2}{#2}}}\frontmatter@RRAPformat}
  {}{}
}%
\makeatother

\begin{document}

This article may be downloaded for personal use only. 
Any other use requires prior permission of the author and AIP Publishing. 
This article appeared as Stolp et al. Rev. Sci. Instrum. 96, 083707 (2025) and may be found at https://doi.org/10.1063/5.0274044 Copyright 2025 Author(s). 
This article is distributed under a Creative Commons Attribution-NonCommercial-NoDerivs 4.0 International (CC BY- NC-ND) License.
\newpage

\preprint{AIP/123-QED}

\title{Optics for broadband x-ray ptychography}

\author{Wiebe Stolp}
\email{wiebe.stolp@ugent.be}
\affiliation{UGCT-RP, Department of Physics and Astronomy, Ghent University, Ghent 9000, Belgium}

\author{Silvia Cipiccia}
\affiliation{Department of Medical Physics and Biomedical Engineering, University College London, London WC1E 6BT, United Kingdom}

\author{Darren Batey}
\affiliation{Diamond Light Source, Harwell Science and Innovation Campus, Fermi Avenue, Didcot OX11 0DE, United Kingdom}

\author{Matthieu Boone}
\affiliation{UGCT-RP, Department of Physics and Astronomy, Ghent University, Ghent 9000, Belgium}

\newcommand{\comment}[1]{\color{red}(#1) \color{black}}
\date{\today}

\begin{abstract}
In conventional x-ray ptychography, diffraction data is collected by scanning a sample through a monochromatic, and spatially coherent, x-ray beam. A high-resolution image is then retrieved using an iterative algorithm. Combined with a scan of the incident photon energy, it is also possible to access chemical and elemental information. Although powerful, the high brilliance required currently constrains the method to 3rd and 4th generation synchrotron sources and long scanning times. An alternative approach is to use a broadband illumination in combination with an energy resolving detector. These detectors record the data in a series of energy channels simultaneously, creating a stack of coherent data suitable for a ptychographic reconstruction. This approach promises to unlock the full power of the radiation source and provide spectral imaging at a higher rate and in a single acquisition. However, these detectors currently saturate well below reaching the flux rates produced at synchrotrons, which is preventing the uptake of this approach. Furthermore, current monochromatic synchrotron setups typically employ Fresnel zone plates for pre-sample focusing due to their stability, flexibility, and affordability, but these diffractive optics limit the spectral bandwidth that the setup can accept. In this article, we analyze the problem and consider alternative optics that can both maximize the total photon detection rates and broaden the tolerable bandwidth. Broadband x-ray ptychography has the potential to dramatically reduce data collection times at synchrotron sources, but also to harness the full power of lower brilliance sources and transition x-ray ptychography into a laboratory technique.
\end{abstract}

\maketitle

\section{Introduction}
\begin{figure}
    \centering
    \includegraphics[width=\linewidth]{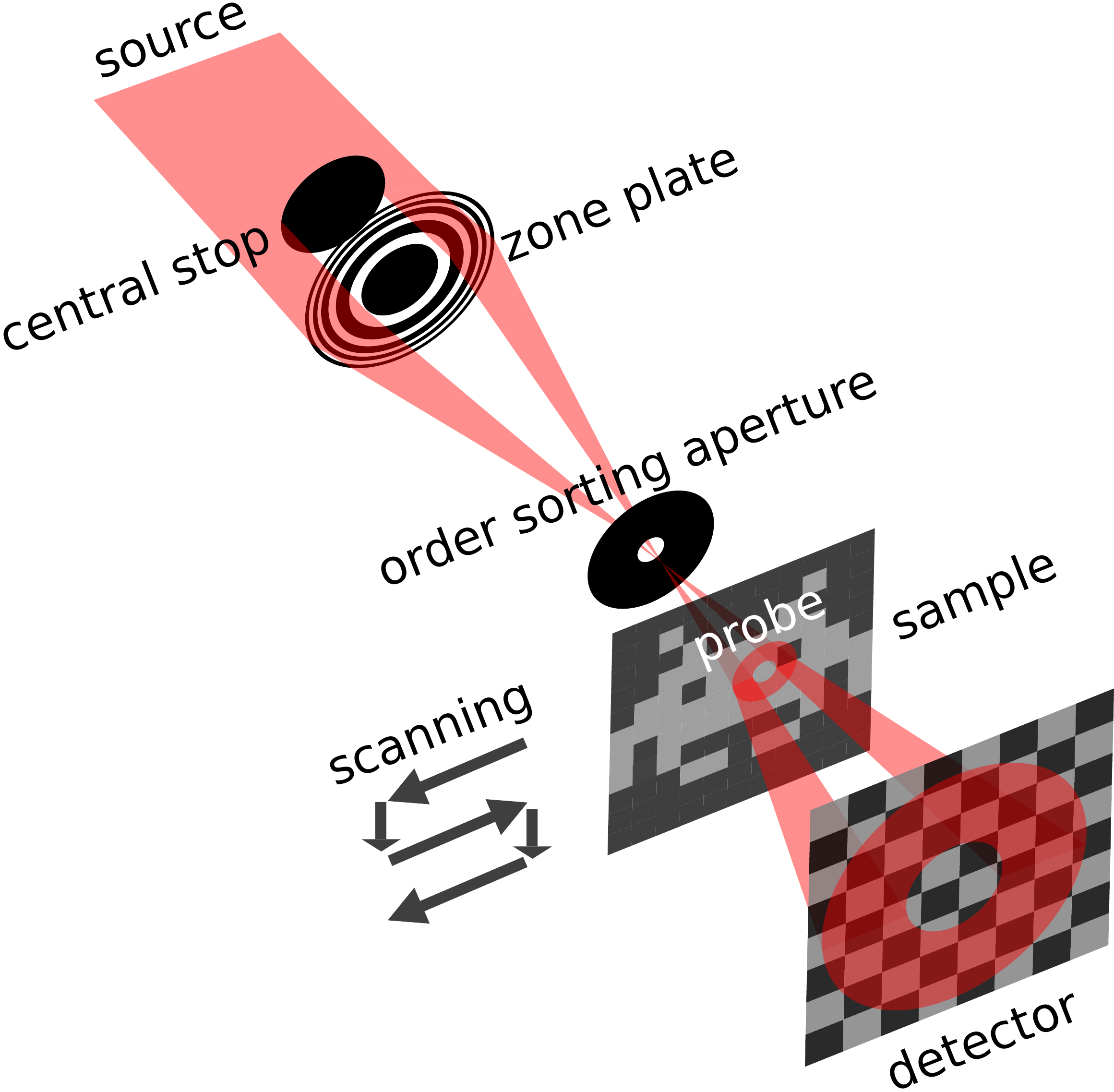}
    \caption{Illustration of the ptychographic acquisition process}
    \label{fig:setup}
\end{figure}

Ptychography \cite{rodenburgHardXRayLenslessImaging2007} is a 2D imaging technique in which a coherent illumination, the probe, is step-scanned through a sample while collecting diffraction patterns (see figure \ref{fig:setup}), with probes at neighboring positions having a degree of overlap. 
After acquisition, the diffraction patterns are processed with an iterative computer algorithm that, by solving the phase problem, retrieves a complex image in which the modulus represents the sample's local transmission and the phase is proportional to the electron density. 
The technique is theoretically diffraction limited and nanometer resolution has been shown to be achievable with x-rays \cite{aidukasHighperformance4nmresolutionXray2024}.
In addition, high-resolution ptychographic reconstructions from data taken at different x-ray energies can be used for material identification by means of spectroscopy \cite{beckersChemicalContrastSoft2011, maidenSoftXraySpectromicroscopy2013, millePtychographyCarbonKedge2022}.
Such spectral ptychographic data has conventionally been acquired sequentially, by adjusting the x-ray monochromator energy and repeating the acquisition, which significantly increases the data collection time. 

By filtering out most of the spectrum, the x-ray monochromator also provides the required temporal coherence. 
X-ray ptychography is conventionally only done with synchrotron sources, which are sufficiently brilliant to perform scans within reasonable time frames despite the remarkable reduction in flux due to the spectral filtering.
However, it has been proven \cite{bateySpectroscopicImagingSingle2019} that a hyperspectral detector can be used instead of a monochromator to provide a sufficient degree of temporal coherence, while also acquiring data over a range of different energies simultaneously. 
By resolving the energy of individual photons, these hyperspectral detectors enable detected photons to be grouped into a large number of energy bins in post-processing.
This process generates a set of co-registered datasets with narrow, adjustable bandwidths, each of which can be independently processed using the ptychographic reconstruction algorithm.
However, the advantages of these detectors are constrained by the particularly low maximum flux tolerated by this type of detector. 
For instance, the SLcam detector \cite{ordavoNewPnCCDbasedColor2011} used in \cite{bateySpectroscopicImagingSingle2019} has a flux limit of approximately 5 photons/pixel/second.
At synchrotrons, where much higher flux levels are available, this limitation undermines the potential efficiency gains that could result from eliminating the monochromator.
In contrast, laboratory x-ray sources, with their naturally lower flux and greater accessibility, would appear better suited to leveraging the efficiency gains from removing the monochromator and employing a hyperspectral detector.
While `multiple peak' spectral laboratory ptychography has already been demonstrated in the visible light regime \cite{loetgeringAdvancesLaboratoryscalePtychography2022, zhangPtychographicHyperspectralSpectromicroscopy2016}, conventional laboratory x-ray sources cannot provide enough spatial coherence for ptychography due to their low brilliance \cite{dukeSynchrotronRadiationProduction2000}.

The development of more brilliant laboratory x-ray sources \cite{larsson24KeVLiquidmetaljet2011} can mitigate this problem, and has already enabled the first lab based ptychography experiment with a hyperspectral detector \cite{bateyXRayPtychographyLaboratory2021}.
In addition to the spectral imaging capabilities introduced by the detector, a larger part of the source flux can be put to use in this configuration, which brings down acquisition time.
This is because instead of the fixed and highly monochromatic filtering of the monochromator, there is now the possibility to filter the spectrum down to a flexible bandwidth in post-processing, which affects the final reconstruction in different ways.
Apart from the energy resolution, the energy bandwidth also affects the spatial resolution of a ptychographic reconstruction, i.e. an increased bandwidth can improve the spatial resolution through photon statistics, but also degrade it through the decreased temporal coherence \cite{spenceCoherenceSamplingRequirements2004}.

Beside the detector, a second key element for broadband ptychography is the pre-sample focusing optics.
The typical x-ray ptychography setup employs a Fresnel zone plate (FZP) for this, but FZPs are diffractive optics and the chromaticity inherent in the diffraction process causes the probe size to vary drastically across energies.
This ultimately increases the total scan duration, as explained in sections \ref{sec:fzp_mono_ptycho} and \ref{sec:fzp_chromaticity}.
This effect has, until recently \cite{stolpBroadbandPtychotomographyHyperspectral2025}, hindered the possibility for hyperspectral ptychotomography.
In this manuscript we explore some strategies for the optimization of a broadband ptychography setup, in particular the pre-sample optics.
As there is no single best solution, and trade-offs need to be made, this work should be read as a comparative overview of the corresponding considerations, differences, strengths and weaknesses.

\section{Fresnel zone plates in monochromatic ptychography} \label{sec:fzp_mono_ptycho}
The focal length at a wavelength $\lambda$ for diffraction order $n \in 2 \mathbb{Z} - 1$, of an FZP with a diameter $D_\mathrm{FZP}$ and an outer zone width $\Delta_\mathrm{FZP}$, is \cite{paganinCoherentXRayOptics2006}:
\begin{equation} \label{eq:focal_length}
    f_\lambda = \frac{D_\mathrm{FZP} \Delta_\mathrm{FZP}}{n \lambda},
\end{equation}
Unwanted diffraction orders, usually $n \neq 1$, are blocked from illuminating the sample using a combination of a central stop close to the FZP, and an order sorting aperture (OSA) in its focal plane as illustrated in figure \ref{fig:setup}.

Equation \ref{eq:focal_length} translates to a formula for the geometrical probe size in the parallel sample plane a distance $z$ downstream of the FZP,
\begin{equation} \label{eq:probe_size}
    D_\lambda(z) = \left| \frac{n z \lambda}{\Delta_\mathrm{FZP}} - D_\mathrm{FZP} \right|,
\end{equation}
where $|\cdot|$ means taking the absolute value.
Equation \ref{eq:probe_size} is also valid at larger distances and can thus be used to calculate the size of the direct illumination disc in the detector plane.

The converging of the wavefield created by the FZP has three benefits:
\begin{enumerate}
    \item \label{point:probe_control}
    Probe size control.
    In the sampling theory of coherent diffractive imaging, the diameter of the probe at the sample is an important parameter \cite{edoSamplingXrayPtychography2013, spenceCoherenceSamplingRequirements2004} since larger probes allow for a larger step size in the ptychographic scan, but also cause smaller fringes in the diffraction patterns in the detector plane.
    These fringes require finer sampling, which constrains the experimental geometry.
    More precisely, to avoid aliasing of the diffraction signal when it is sampled by the detector we must satisfy the coherent diffraction imaging sampling condition \cite{spenceCoherenceSamplingRequirements2004}, which states that the actual probe diameter $D_\lambda(z_\mathrm{obj})$ in the object plane must not be larger than the maximum diameter $D_\lambda^*$ dictated by the Nyquist-Shannon sampling condition, i.e.
    \begin{equation} \label{eq:sampling}
        D_\lambda(z_\mathrm{obj}) \leq D_\lambda^* = \frac{Z_\mathrm{det}}{2\Delta_\mathrm{det}} \lambda ,
    \end{equation}
    Here $\Delta_\mathrm{det}$ is the pixel pitch of the detector and $Z_\mathrm{det}$ its distance from the sample. 
    Looking at equation \ref{eq:probe_size}, this requirement can be easily satisfied by changing $z_\mathrm{obj}$, i.e. by moving the sample plane along the optical axis.
    The maximum scan step size for a given probe overlap can only be achieved when $D_\lambda(z_\mathrm{obj}) = D_\lambda^*$.
    \item \label{point:beam_spreading}
    Detector plane beam spreading. 
    The FZP produces a converging beam that transitions into a diverging beam beyond focus, forming a relatively uniform, disc-shaped intensity distribution in the far field. 
    When the numerical aperture of the FZP is sufficiently large and the detector is sufficiently far, this uniform disc occupies a significant portion of the detector's active area. 
    Since the entire disc encodes information about the sample, this configuration ensures efficient utilization of both the detector and the x-ray beam. 
    Such efficiency is especially critical for detectors with low count-rate capabilities, such as hyperspectral detectors.
    \item \label{point:high_spatial_frequencies}
    High spatial frequencies. The diverging probe wavefront contains (by its very nature) high spatial frequencies, which has been shown to benefit reconstruction quality \cite{guizar-sicairosRoleIlluminationSpatialfrequency2012}. This is different from point \ref{point:probe_control}, which concerned the spatial frequencies contained in the far field diffraction pattern.
    \end{enumerate}

\section{Fresnel zone plate chromaticity in broadband ptychography} \label{sec:fzp_chromaticity}
In the case of a moderately broad spectrum, far field aliasing can still be avoided by moving the sample closer to the FZP foci to reduce $D_\lambda$ for all $\lambda$.
However, if the relevant spectrum is too broad for a given FZP, no sample plane exists in which this equation is satisfied by all energy-probes. 
Furthermore, even if such a plane \emph{does} exist, the smallest probe in the spectrum dictates the appropriate scan step size due to the required probe overlap.
This step size will be significantly smaller than what would be necessary for the largest probe, meaning that more steps will always be required for a broadband acquisition than for a monochromatic acquisition.
Although this is not a fundamental issue - by reducing the exposure time of each step one can keep the total time spent on data acquisition constant - the overhead that exists with each step increases the scan duration in practice.
While the ptychographic reconstruction algorithm can be adapted to allow for over-sized probes and even remove the probe diameter constraint altogether \cite{bateyReciprocalspaceUpsamplingRealspace2014}, this adaptation relies on the additional redundancy due to the increased overlap between such probes.
In other words, an increased probe size does not allow an increased step size in this scenario, and therefore this adaptation can not reduce the total number of steps.
In addition to the added overhead, an increased number of steps also corresponds to a larger total data size, thereby placing greater demands on detector throughput and data storage requirements, and adding strain to the ptychographic reconstruction algorithm.

Apart from increasing the number of steps, probes that vary too much with energy also interfere with the reconstruction in another way.
The conventional ptychographic model is based on the propagation of a single wave field, but the energy bins of a spectral acquisition must have a certain finite width to contain sufficient photons, and therefore each bin always contains a range of wavefields.
Even if sufficient photons are a given, a very narrowly defined energy bin will still contain multiple energy wavefields outside its defined band due to the finite energy resolution of the detector.
Thus, a probe that is relatively stable within each energy bin, not only in size but in terms of its complete complex function representation, will fit the ptychographic single wavefield model better than one that varies strongly.
A better fit finally increases the likelihood of a successful reconstruction.
One could adapt the model by accounting for the different energies within a bin in the algorithm \cite{bateyInformationMultiplexingPtychography2014}, but this requires additional photons.

The energy dependent focal length also creates issues with the OSA, the pinhole used to block the FZP's unwanted diffraction orders from hitting the sample.
These can be easily solved by making sure the central stop is larger than the OSA, but this will decrease the surface area of the FZP that is effectively used.

In the following, the collection of problems resulting from the overly strong probe energy dependence is called the probe size problem.
Combined with the low count rate of hyperspectral detectors, this problem has motivated us to investigate the strategies described in sections \ref{subsec:pin_BS}-\ref{subsec:szp}.

\section{Optics strategies for broadband ptychography} \label{sec:strategies}
\subsection{Achromatic optics and related work} \label{subsec:achromatic}
A natural choice to get the benefits of focusing optics without the chromaticity problems of the FZP is the use of achromatic focusing optics, e.g. a pair of Kirkpatrick–Baez focusing mirrors. 
However, these are relatively expensive, especially in the context of a laboratory setup.
Furthermore, perfect achromaticity is actually not ideal in real-world spectroscopic ptychographic imaging.
Firstly, as seen in equation \ref{eq:sampling}, the ideal probe size \emph{is} chromatic since it scales with $\lambda$. 
It should just not scale much faster than that, as is the case with a (large) FZP.
Secondly, photons over all energies count towards the upper flux limit of the hyperspectral detector, while in most cases, not the whole spectrum is of interest for the experiment.
Indeed, the same FZP chromaticity that has been described as a problem earlier, also works as a beneficial spectral filter.

A recently developed optics combines a compound refractive lens and an FZP which counterbalance each other’s chromatic effects \cite{kubecAchromaticXrayLens2022}. 
While promising, this design is limited to a specific energy range, making it inflexible, and is furthermore costly and not yet readily available. 
Another interesting solution is an FZP with a zone profile that is algorithmically optimized to reduce the chromatic focusing effect \cite{loetgeringTailoringSpatialEntropy2021}), but we became aware of this design only recently.

\subsection{Pinhole with beam stop or attenuator in far field} \label{subsec:pin_BS} 
Chromaticity might also be avoided by using a pinhole to define the probe, which is significantly less costly than KB mirrors or other advanced optics.
The downside of this approach is the lack of divergence in the beam (point \ref{point:beam_spreading} in section \ref{sec:fzp_mono_ptycho}). 
More specifically, a probe that is defined by a pinhole is locally similar to a plane wave and thus generates far-field diffraction patterns similar to a delta function, i.e. with a high dynamic range - intensely bright at the center but with low intensity at the outer regions. 
This uneven distribution is challenging to record with most detectors, but especially so with the hyperspectral cameras considered in this work.
This is a well known problem that can be, and often is, mitigated by blocking the center region of the diffraction pattern with a beam stop just before the detector.
In so far as information about a particular sample frequency is shared between pixels, obstructed pixel values can be recovered from unobstructed pixels during the ptychographic reconstruction, using a technique called ``detector pixel floating'' \cite{rodenburgHardXRayLenslessImaging2007, maidenSuperresolutionImagingPtychography2011}.
However, a beam that approaches a delta function in the far field localizes sample frequency information to an increasingly small region, covering only a small number of neighboring detector pixels or even just a single one.
If the beam stop is too large, there is thus insufficient information available in unobstructed pixels, hindering recovery of the values in the obstructed ones. 
On the other hand, assuming the pinhole diffraction pattern intensity profiles roughly scale with the inverse distance from the optical axis, one will need a large beam stop to get to a satisfactory dynamic range.
In other words, as the beam stop increases in size, the patterns can be more efficiently recorded due to the lower dynamic range, but will be more difficult to reconstruct using the pixel floating method due to the increasing amount of unrecoverable information.
We have found, presumably for this reason, that when using the pinhole setup with the SLcam hyperspectral detector, the effective acquisition rate was still too low to produce ptychographic reconstructions reliable enough for spectroscopic imaging in an acceptable time frame.

A potential approach to better mitigate the uneven flux distribution is to use a beam \emph{attenuator} positioned directly before the detector, to attenuate rather than completely block the beam.
Ideally, such an attenuator has a variable thickness that ``reverses'' the intensity distribution of the pinhole diffraction pattern, i.e. it should make the distribution more homogeneous.
Assuming the transmission function of the attenuator is known, the recorded diffraction patterns can then be corrected before doing the reconstruction by dividing out the attenuator transmission.
A two level prototype of such a variable thickness attenuator was created by combining a small hard (full) attenuator in the optical axis with a thinner, soft (partial) attenuator surrounding it.
\begin{figure}
    \centering
    \includegraphics[width=\linewidth]{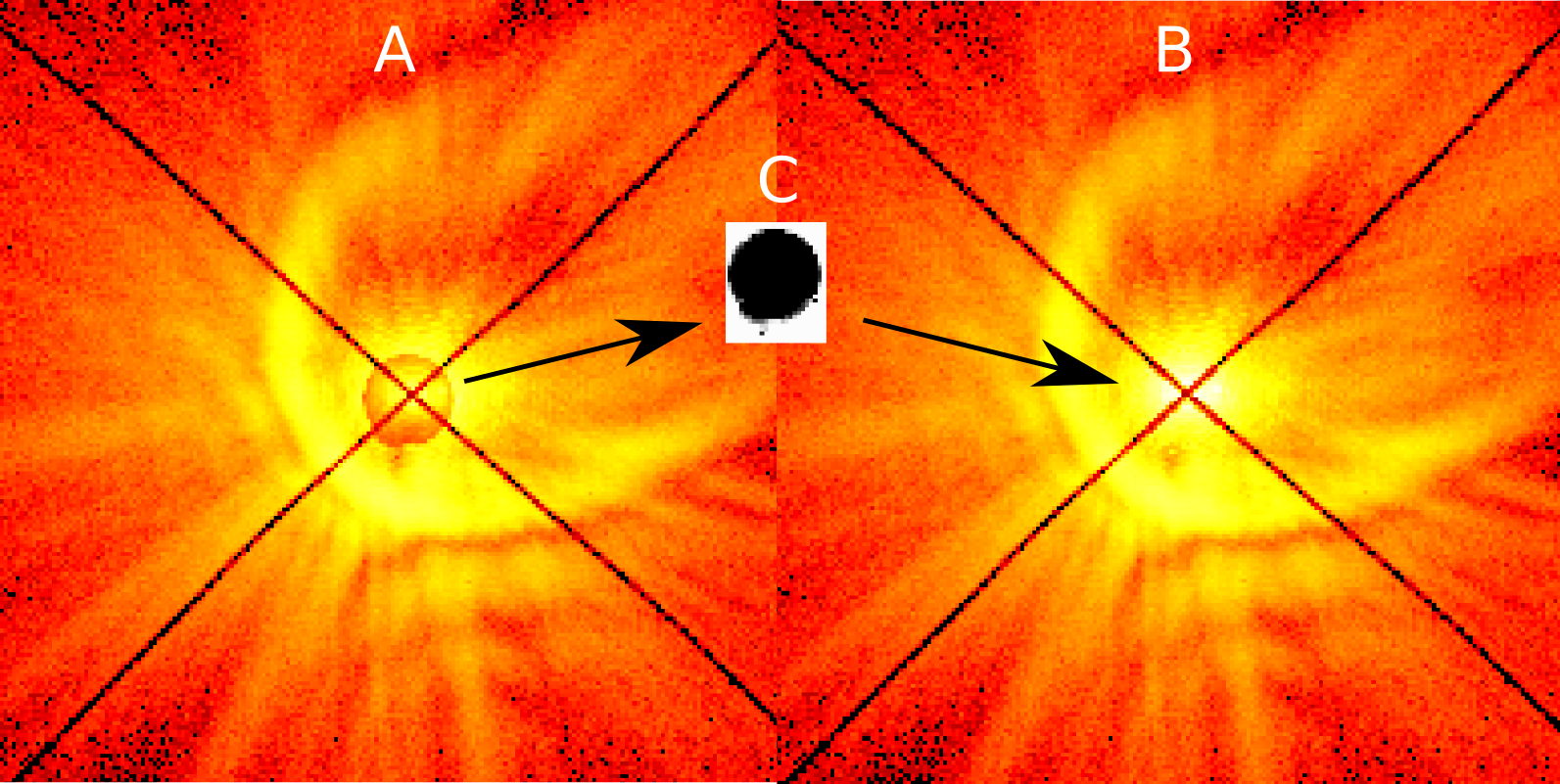}
    \caption{Illustration of the diffraction pattern correction process.
    A shows, using a log scale, the sum of all diffraction patterns of a ptychographic scan of a Siemens star test sample, within a narrow energy band (8.0 $\pm$ 0.1 keV).
    B shows the same thing after correcting for the soft attenuator visible in the center of A. 
    The thickness of the soft attenuator is shown in C, exposing partial volume effects at its edge.
    The diagonal cross visible in both A and B is part of the orignal acquisition. 
    It results from the hard attenuator that serves only to block the brightest part of the diffraction pattern in the center. 
    No manual correction is applied in this cross region.
    }
    \label{fig:attenuator_correction_illustrated}
\end{figure}
The soft attenuator was made of $\sim$7 µm thick gold foil, approximately circular with a diameter of $\sim$1.0 mm.
The hard attenuator was constructed by crossing two 100 µm diameter gold wires in the optical axis.
This two level attenuator was used in an experiment at the Coherence branch of beamline I13 at the Diamond Light Source.
A broadband spectrum was created by tapering the undulator \cite{mossessianCharacterizationALSUndulator1995} and spectral data was acquired using the SLcam hyperspectral detector \cite{bjeoumikhovSLcamFullfieldEnergy2012}.
The two level attenuator was attached to a Kapton foil a few mm upstream of the detector.
A 10 µm pinhole just upstream of the sample served as the pre-sample beam-defining optics.
A Siemens star test sample was scanned in a ptychographic acquisition with a 1 µm step size in a 20x20 grid.
The exact thickness and energy dependent transmission of the soft attenuator were characterized by analyzing diffraction patterns with and without the two level attenuator.
Figure \ref{fig:attenuator_correction_illustrated} illustrates the correction process, with diffraction data covering a nominal energy range of 8.0 $\pm$ 0.1 keV.

\begin{figure*}
    \centering
    \includegraphics[width=\linewidth]{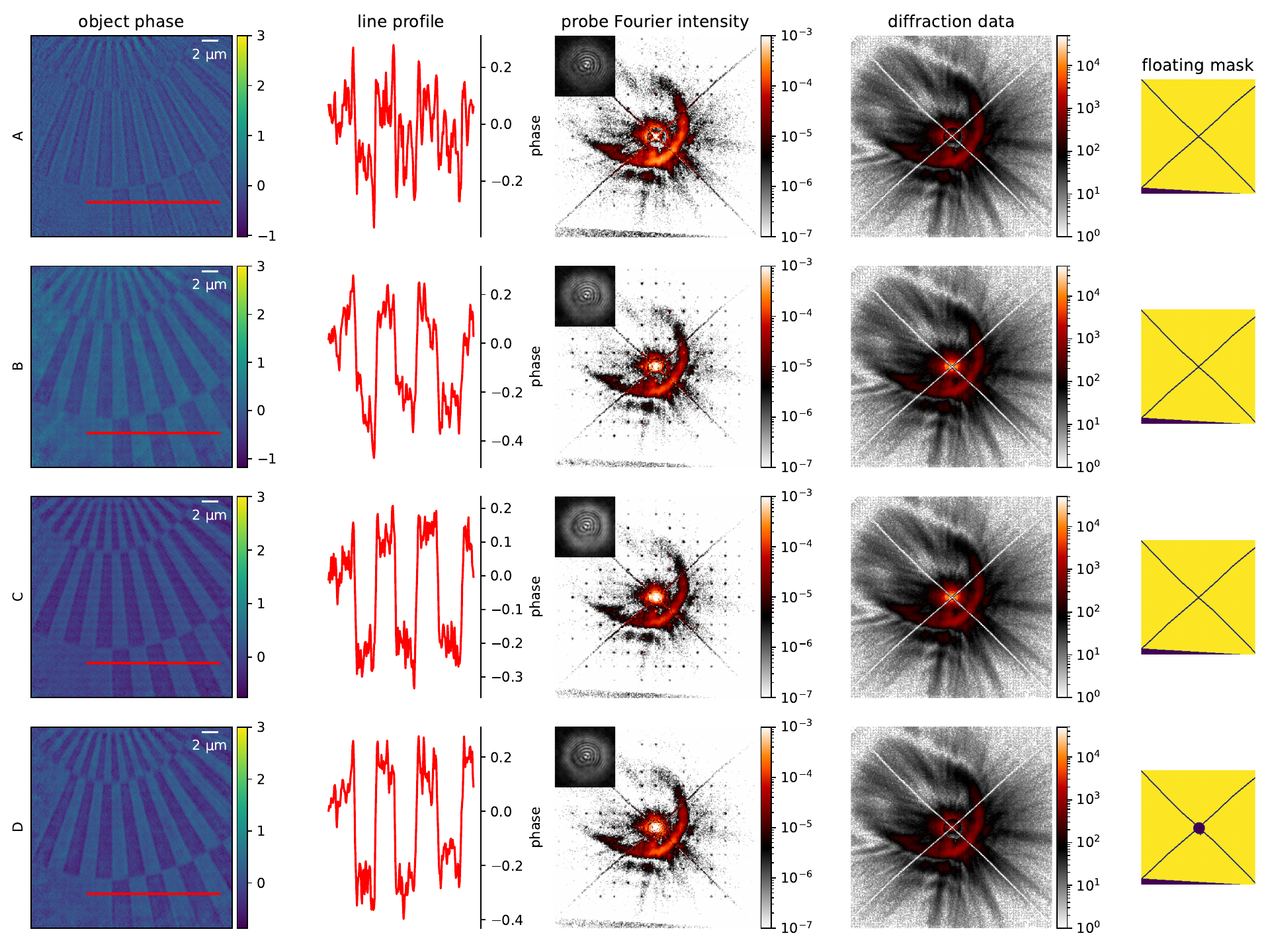}
    \caption{
    Different ways to correct for a circular soft beam attenuator in the optical axis in a ptychographic acquisition of a Siemens star test sample. 
    Phase reconstructions are shown in the \emph{object phase} column.
    For an additional way to compare the phase reconstructions, the \emph{line profile} column shows the mean phase along a 10 pixel wide line, crossing three spokes of the Siemens star.
    The \emph{probe Fourier intensity} column shows the simulated far field intensity of the recovered probe in the absence of a sample and furthermore contains the real space probe moduli in the insets, normalized to have equal total intensity across rows.
    \emph{Diffraction data} shows the summed intensity of all diffraction patterns that were used as input for the reconstruction.
    \emph{Floating mask} shows the binary mask of pixels set to float during the reconstruction. 
    Different rows correspond to different correction methods:\\
    A: No correction applied at all. Diffraction data contains photon energies within 8.0 $\pm$ 0.1 keV.
    B: Manual correction applied to the same diffraction data, as illustrated in figure \ref{fig:attenuator_correction_illustrated}.
    C: Manual correction using two different energy bandwidths to improve statistics: 8.0 $\pm$ 0.5 keV behind the soft attenuator and 8.0 $\pm$ 0.1 keV outside of it.
    D: Without manual correction, but with automatic algorithmic pixel floating correction during the reconstruction. Diffraction data within 8.0 $\pm$ 0.1 keV.
    A version of this figure that includes the object modulus, with its own line profile, is available in the supplementary material.
    }
    \label{fig:attenuator_correction_comparison}
\end{figure*}

A more detailed analysis of the same data is shown in figure \ref{fig:attenuator_correction_comparison}.
This figure compares three different types of diffraction pattern correction and the effect on the corresponding reconstructions.
Reconstructions were done using an implementation of the ePIE algorithm \cite{maidenImprovedPtychographicalPhase2009} which was run for 1000 iterations, each reconstruction taking approximately 70 minutes on an Nvidia Geforce GTX 1080 Ti GPU. 
Pixel floating was employed behind the hard attenuator in all reconstructions.

Figure \ref{fig:attenuator_correction_comparison}A shows an uncorrected dataset composed of photons with nominal energies of 8.0 $\pm$ 0.1 keV, and the corresponding reconstruction.

In figure \ref{fig:attenuator_correction_comparison}B, the same selection of data was used, but in this case the data was manually corrected before reconstruction.
This was done by dividing out the energy dependent soft attenuator transmission profile from 50 eV binned spectral diffraction data and then summing four energy bins to obtain a dataset with a nominal 200 eV bandwidth.
The reconstructions in this row contain low frequency artifacts, which is likely a pitfall of the correction process: the high photon shot noise corresponding to the low intensity measurements behind the soft attenuator is transferred to the corrected intensity values, resulting in the unnatural scenario of high intensity measurements with a high degree of uncertainty.

In figure \ref{fig:attenuator_correction_comparison}C, a way to improve on this is investigated. 
From the same spectral dataset, a selection was made with mixed energy bandwidths across the different pixels, i.e. a 1 keV bandwidths was used behind the soft attenuator, while the same 200 eV was used outside of that. 
Appropriately mixing the two bandwidths into a single diffraction pattern was done as follows: 
first, the same energy dependent correction as in figure \ref{fig:attenuator_correction_comparison}B was performed on the 50 eV binned dataset. 
From the corrected dataset, 20 bins were summed to obtain a dataset with a 1 keV wide energy bandwidth, and four bins were again summed to obtain the 200 eV bandwidth dataset. 
Each individual full dataset was then normalized to have a total intensity equal to 1 across all scan steps together, allowing the pixel values to be interpreted as probabilities in a hyper-dimensional space.
Then, the two energy ranges were mixed, taking values from the broad bandwidth dataset behind the attenuator and from the narrow bandwidth dataset outside of that.
Lastly, the whole mixed energy dataset was multiplied with a certain constant that preserves the original photon counts in the pixels coming from the narrow bandwidth dataset, which was only necessary to be able to compare the diffraction data to that of the other methods in figure \ref{fig:attenuator_correction_comparison}.

This mixing method is defensible, since for any fixed bandwidth, lower sample frequencies subtend a smaller angle on the detector than higher ones.
In other words, lower frequency information is less energy dependent and thus suffers less from a summation across energy.
A more sophisticated version of this principle could be done, where the energy bandwidth gradually increases towards the optical axis, regardless of the presence of any attenuator.
However, this method only makes sense if we can assume our probe to be similar to a plane wave like in this dataset, since only then is the information regarding sample frequencies spatially organized in this manner.

Finally, in figure \ref{fig:attenuator_correction_comparison}D, uncorrected 200 eV bandwidth data is again used, but this time the pixel floating correction method was also used in the pixels behind the soft attenuator, in addition to the ones behind the hard attenuator.
Note that while the reconstruction of \ref{fig:attenuator_correction_comparison}C has access to more sample information than that of \ref{fig:attenuator_correction_comparison}A and \ref{fig:attenuator_correction_comparison}B, the one of \ref{fig:attenuator_correction_comparison}D has access to less, since the information behind the soft attenuator is discarded. 
One must therefore be careful in comparing them.

The pixel floating reconstruction algorithm tended to fill the masked pixels behind the hard attenuator cross with unrealistic values, as can be seen in the far field probe column.
This is more prominent and also more expected in case of the uncorrected data of figure \ref{fig:attenuator_correction_comparison}A, where the missing intensity is likely being placed there where it conflicts least with the model.
However, in the three other (milder) cases of figures \ref{fig:attenuator_correction_comparison}B-D this behavior is somewhat unexpected.
Furthermore, all far field probes contain a grid artifact, which is worse in the two manual correction methods of figures \ref{fig:attenuator_correction_comparison}B and D.
To varying degrees, the scenarios of figures \ref{fig:attenuator_correction_comparison}B-D also suffer from another grid artifact, namely in the object reconstruction, most prevalent in the manual mixed energy correction of figure \ref{fig:attenuator_correction_comparison}C.
However, that reconstruction does have an improved low frequency noise compared to \ref{fig:attenuator_correction_comparison}B and D.
The line profiles show an improvement going from A to B to C, with D being roughly comparable to C.
Position correction \cite{maidenAnnealingAlgorithmCorrect2012} was not used in any of the reconstructions due to instability with this dataset.

In conclusion, while the manual correction methods mostly improve the reconstruction compared to not doing a correction at all, manual correction gives results somewhat comparable to the automatic pixel floating method with this Siemens star dataset, even though that method had less information to its disposal and is functionally equivalent to using a hard beam stop.
Similar to reconstructing acquisitions that actually used such a hard beam stop, with the soft attenuator we did not manage to get a satisfactory modulus signal when reconstructing more realistic samples.
However, the modulus signal is crucial for spectral imaging.
This problem is also visible in the modulus signals in figure \ref{fig:attenuator_correction_comparison}, although one cannot easily generalize between different types of samples.
The fact that pixel floating results in similar results as manual correction seems to imply that the diffraction patterns are still too underexposed behind the attenuator, and therefore cannot be properly corrected even with perfect knowledge of the system.
This would also mean that an attenuator that better matches the distribution of the pinhole diffraction pattern could outperform the pixel floating method, but such an attenuator is challenging to produce, would be tied to the particular pinhole in question, and also to its orientation.
Furthermore, it might still not suffice when imaging strongly scattering samples that significantly change the intensity distribution of the diffraction pattern produced by the pinhole on its own.
Another drawback of this method is that, even if the acquired dataset yields good reconstructions, a significantly higher portion of the flux passes through the sample than was recorded, potentially resulting in unnecessary sample damage.
Lastly, even though the attenuator was placed close to the detector, diffraction effects might influence the intensity distribution on the detector near the edges of the attenuator, which cannot be corrected for using a mere division.
It should be noted that this experiment was influenced by the presence of the hard attenuator outside of the optical axis, where it served no purpose other than support.
Finally, while it is conceivable that a better tuned version of this setup could produce satisfactory results, it is too far from a simple solution for us to have explored it any further.

\subsection{Pinhole with diffuser}
In the context of this article, based on previous considerations and experimental tests, an ideal pre-sample focusing optic would spread the beam across the entire detector in the far field while keeping it sufficiently small in the sample plane to satisfy sampling equation \ref{eq:sampling} for all energies. 
Simultaneously, it should not make the beam excessively small in order to avoid requiring a smaller scan step size.

Using a pinhole to constrain the beam, combined with a diffuser - a randomly structured material that scatters and spreads the beam - emerges as a natural choice.
Note that the current focus is the diffuser's beam spreading effect on the detector, not the creation of a highly structured illumination which can also aid ptychographic reconstruction in different ways \cite{odstrcilOptimizedIlluminationHighresolution2019a, guizar-sicairosRoleIlluminationSpatialfrequency2012}.
By the convolution theorem, the far field diffraction pattern of any object is the Fourier transform of its autocorrelation function. 
If we assume this object autocorrelation function to be gaussian with a full width at half maximum (FWHM) $a$, its diffraction pattern is also gaussian, with FWHM $\lambda z / a$.
Thus with 0.15 nm wavelength radiation (8.3 keV) we would need a sample plane autocorrelation FWHM of around 300 nm to fill a 1 cm wide detector with the FWHM in that plane after 10 meters propagation.
Finding a material with such an autocorrelation function that does not absorb too much of the intensity is not trivial.
However, if the beam spreading effect of the diffuser is insufficient, i.e. the center spot in the far field is too bright, a beam stop may again be used before the detector.
This diffuser-beam stop combination can still offer an advantage over the beam stop on its own, since with this combination the low frequency information of the sample is spread out to illuminate the detector beyond the beam stop shadow.
Since this causes information about any particular sample frequency to be shared over a larger amount of pixels in the diffraction pattern, lost parts of the pattern, e.g. due to the beam stop, can be recovered more easily by the pixel floating algorithm.
This finally reduces the exposure time needed to collect enough photons for each diffraction angle and get to a desired reconstruction quality.

We have analyzed three random diffuser materials in combination with a 5 µm pinhole:
\begin{itemize}
    \item 10 µm tall disordered silicon nanowires of ca. 100 nm diameter on a 500 µm thick Si substrate. While this is very thick, the current investigation was centered around the beam spreading effect and not the absorption.
    \item Iridium-coated nanowires, identical in structure to the first sample but supplemented with an atomic layer deposition of iridium.
    \item A dried solution of 100 nm Au nanospheres on Kapton.
\end{itemize}

\begin{figure*}
    \centering
    \includegraphics[width=\linewidth]{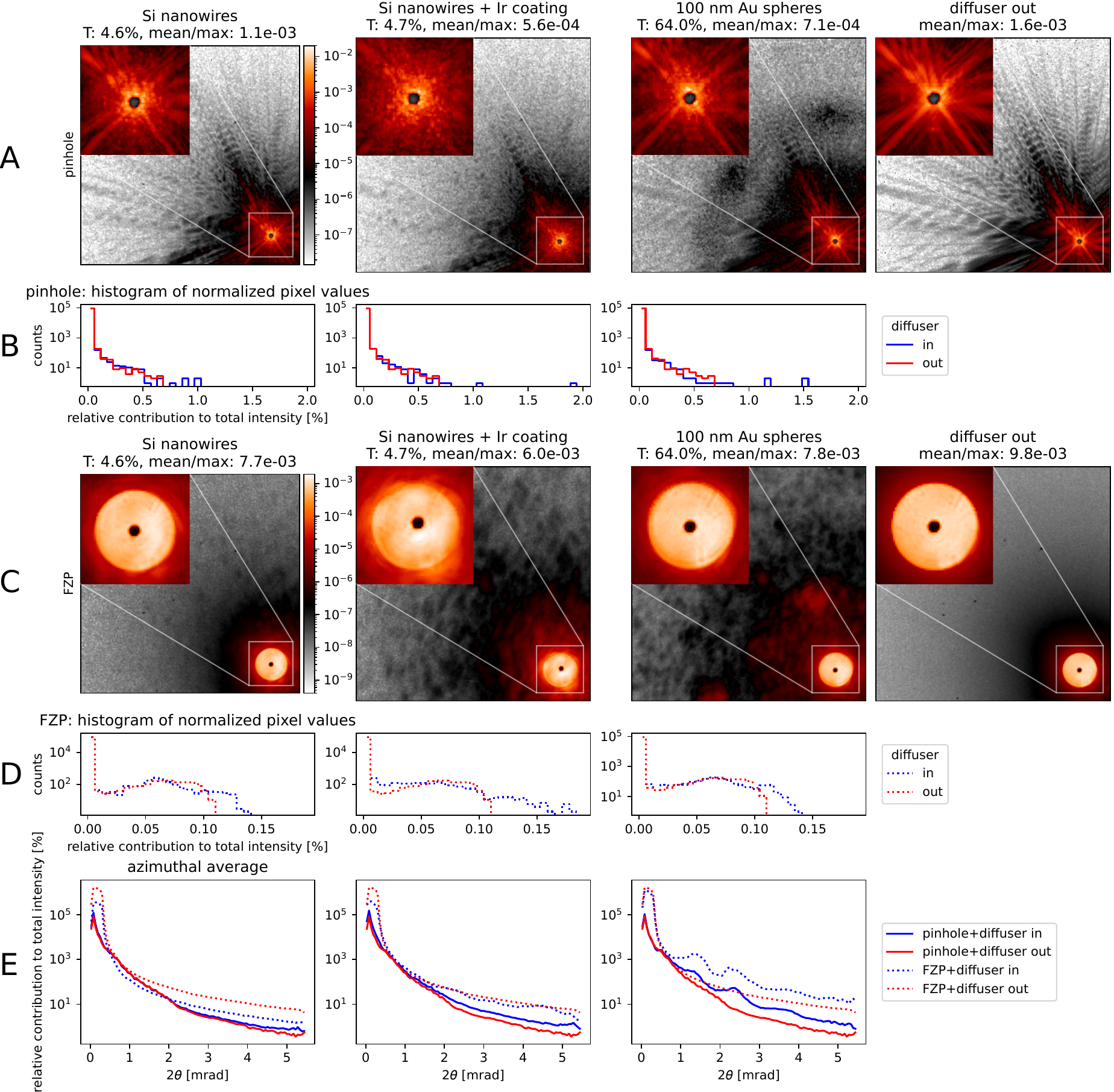}
    \caption{Analysis of diffraction patterns of three diffusers, using either a pinhole or an FZP as optics upstream of the diffuser.
    The last column contains the images without any diffuser for reference.
    All images are created by first summing and then normalizing all diffraction patterns of a ptychographic scan of a Siemens star test sample (not shown). 
    Zoomed-in insets are included to highlight the details of the images near the optical axis.
    T refers to the total transmission of the diffuser.
    A: Using a pinhole as optics. 
    B: Histograms of A, showing that diffusers tend to generate relative hot spots.
    C\&D: Analogous to A\&B, but with an FZP as optics to eliminate the effect of the beam stop on the analysis, and to enhance the visibility of the beam spreading effect.
    E: An azimuthal integration around the optical axis of all images, as a function of diffraction angle $2 \theta$ to visualize the beam spreading effect. 
    }
    \label{fig:diffusers}
\end{figure*}

The results of this analysis are shown in figure \ref{fig:diffusers}.
The x-ray spectrum used for this experiment was monochromatic, centered at 9.7 keV, and the acquisitions were taken with an Eiger photon counting detector.
Despite the use of a diffuser, a 0.6 mm wide detector-side beam stop was still required to reduce the dynamic range for each of the diffusers in the pinhole configuration, which may affect the beam spreading analysis.
To get an extra analysis that is unaffected by a beam stop, the same diffusers were also investigated in an FZP configuration, shown in figure \ref{fig:diffusers} as well.
In this configuration, a 5 µm OSA was placed at the FZP focus, in combination with a 20 µm central stop centimeters upstream of the FZP, as visualized schematically in figure \ref{fig:setup}. 
Note therefore, that while both pinhole and FZP images contain a zero intensity region around the optical axis, the causes of this low intensity are different.

Each of the images in the figure is the sum of all diffraction patterns in identical ptychographic scan trajectories of a Siemens star test sample. 
Due to experimental issues, the data is not suitable for reconstruction, but the diffraction patterns are used here to show the variation in scattering performance.
Since the flux available at synchrotron sources is high compared to the count rate of current hyperspectral detectors, absorption of flux due to the diffuser is not necessarily a limiting factor at this time.
This initial study was therefore purely focused on the beam spreading effect and not on absorption.
To remove the influence of absorption from the data, all images have been normalized by dividing the images, after the mentioned summation, by their own total intensity.
However, the total transmission of the beam by the diffuser, taken from a separate measurement, is still listed above each image.

As a metric to quantify the increased homogeneity of the images in figure \ref{fig:diffusers}A and \ref{fig:diffusers}C, as caused by the beam spreading effect, the ratio between the mean and maximum of each image can be calculated.
This metric approaches 0 when there is no spread at all, and goes to 1 for a homogeneous illumination.
It is not affected by multiplication factors, i.e. changes in total intensity.
The higher this homogeneity metric, the more total flux a detector can capture without saturating any of the pixels, thus facilitating faster acquisitions.
The homogeneity metric is listed as mean/max in figure \ref{fig:diffusers} for each of the displayed diffraction patterns, but it turns out all tested diffusers cause the metric to go down instead of up.
This is because, while diffusers can naturally only reduce the overall intensity, relatively speaking they can introduce hot spots to the diffraction patterns, affecting the metric negatively.
The hot spot effect can be seen more clearly in the histograms below each image (figures \ref{fig:diffusers}B and D), which for reference also includes the histogram of the image produced from patterns with no diffuser.
In the case of the pinhole configuration, hot spots might also be created in the center region due to the direct beam being shifted or broadened beyond the beam stop, but a similar effect is observed in the case of the FZP patterns, where no detector-side beam stop was present.

To make the beam spreading visible in a way that is uncoupled from the hotspot formation, an azimuthal average is calculated around the optic axis for each image using PyFAI \cite{kiefferNewToolsCalibrating2020}, shown in figure \ref{fig:diffusers}E.
This beam spreading effect is more pronounced in the FZP configuration.
In the pinhole configuration, the two Si nanowire diffusers seem to cause the beam to shrink very slightly instead of spread.
However, these graphs are likely again influenced by the beam stop.

Summarizing, the Ir coating on the nanowires has a beneficial effect on the beam spreading, but between all tested diffusers the Au nanospheres appear to be most promising.
However, for each of the tested diffusers the favorable beam spreading effect is overpowered by the unfavorable reduction of the intensity homogeneity, i.e. the creation of hot spots.
The net effect of these diffusers is therefore antithetical to our initial aim.
It is also worth noting that, even if a diffuser would produce a perfect gaussian intensity distribution in the far field, it would still not be as efficient as the homogeneous disc distribution resulting from an FZP.
Of course one could ignore a small amount of hot spots at the expense of a reduced accuracy in the diffraction pattern recordings, but this has unpredictable effects and does not seem worth the possible gains.
Altogether, this might indicate that the randomly structured diffuser type does not fit the proposed requirements.

\subsection{Small diameter Fresnel zone plate} \label{subsec:szp}
Instead of replacing the FZP with a different optics, another solution is to adapt the FZP.
Comparing equation \ref{eq:probe_size} and \ref{eq:sampling}, the ideal $\lambda$ scaling of the probe size is hindered by the FZP diameter $D_\mathrm{FZP}$.
We can therefore get closer to the ideal situation by reducing that parameter.
If we use $z_\lambda$ for the distance at which the probe of wavelength $\lambda$ has the ideal size $D^*_\lambda$ to prevent aliasing in a particular setup (equation \ref{eq:sampling}), then $D_\lambda(z_\lambda) = D_\lambda^*$.
Let $\lambda'$ be another wavelength, which produces another probe with diameter $D_{\lambda^{'}}(z_\lambda)$ in that same plane. 
Then the actual ratio between the two probe sizes is $Q = D_{\lambda'}(z_\lambda) / D_{\lambda}(z_\lambda)$, determined by equation \ref{eq:probe_size}, while the ideal ratio is always $Q^* = \lambda'/\lambda$ according to equation \ref{eq:sampling}.
An error metric to quantify the probe size problem, i.e. the undesired part of the probe size variation over energy, is therefore the absolute difference between these two ratios, which, for any given diffraction order $n$, works out to be
\begin{equation} \label{eq:epsilon}
\begin{split}
	\epsilon = |Q - Q^*| =  \left|\frac{D_{\lambda'}(z_\lambda)} {D_{\lambda}(z_\lambda)} - \frac{\lambda'}{\lambda}\right| & = \left|\frac{D_{\lambda'}(z_\lambda)} {D^*_{\lambda}} - \frac{\lambda'}{\lambda} \right| \\
    & = \frac{|\lambda' - \lambda|} {\lambda} \frac{D_\mathrm{FZP}}{D^*_{\lambda}}
\end{split}
\end{equation}

Evidently, $\epsilon$ does not depend on $z_\lambda$.
Furthermore, it is proportional to the FZP diameter $D_\mathrm{FZP}$ and to the relative bandwidth $|\lambda' - \lambda| / \lambda$.
In other words, if we want to increase the relative bandwidth of our experiment without increasing the probe size variation, we have to decrease the FZP diameter at the same rate.
The metric $\epsilon$ is also inversely proportional to $D^*_\lambda$, but increasing $D^*_\lambda$ can only be achieved through finer far field sampling (equation \ref{eq:sampling}), which requires a larger propagation distance or smaller detector pixels.
It should be noted that in equation \ref{eq:focal_length}, the FZP diameter $D_\mathrm{FZP}$ and outer zone width $\Delta_\mathrm{FZP}$ are treated as independent variables, so in equation \ref{eq:epsilon}, $\epsilon$ does not depend on $\Delta_\mathrm{FZP}$. 

\begin{figure}
    \includegraphics[width=\linewidth]{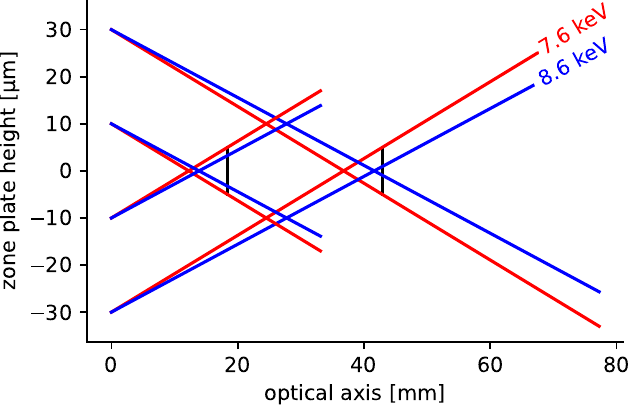}
    \caption{Schematic representation of the probes originating from a 20 µm and a 60 µm FZP at 7.6 and 8.6 keV. For both FZPs, black vertical lines are drawn at the position along the optical axis where the largest probe, i.e. the 7.6 keV probe, is 10 µm wide.}
\label{fig:szp_geometry}
\end{figure}

It is therefore clear that a smaller FZP is able to extend the tolerable bandwidth in broadband ptychography.
Figure \ref{fig:szp_geometry} illustrates this for FZPs with diameters $D_\mathrm{FZP}=20$ µm and $D_\mathrm{FZP}=60$ µm, both with an outer zone width $\Delta_\mathrm{FZP}=100$ nm.
In both cases a black line indicates the position of the sample plane for which a 7.6 keV beam, focused by the FZP, creates a 10 µm probe on the sample.
In that same plane, an 8.6 keV beam now produces a 6.5 µm probe if $D_\mathrm{FZP}=20$ µm, but only a 1.8 µm probe if $D_\mathrm{FZP}=60$ µm. 
Note that the large FZP with $D_\mathrm{FZP}=60$ µm is used for illustration purposes and is still small in comparison with diameters starting from 200 µm used in conventional setups.

\begin{figure*}
    \centering
    \includegraphics[width=\textwidth]{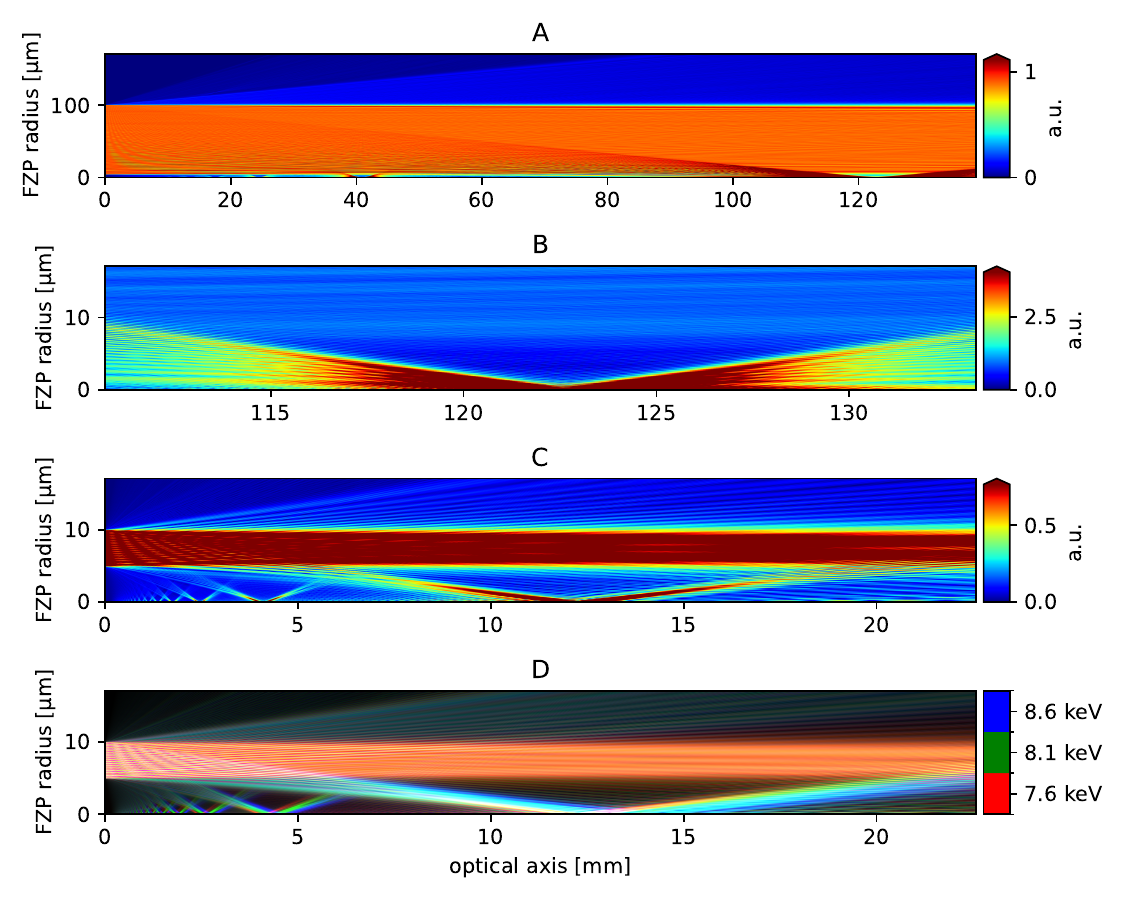}
    \caption{Simulated radial cross-sections of the wavefield magnitudes downstream of two 963 nm thick Fresnel zone plates, performed using PyHank \cite{guizar-sicairosComputationQuasidiscreteHankel2004}.
    Since the fields are radially symmetric, the upper half of the image is not shown. 
    The magnitude of the incoming wavefield is set equal 1 everywhere before it passes through the zone plate, which is embedded in a thick absorbing material.
    A 10 µm diameter, 20 µm thick central stop attenuates the wavefield directly before the zone plates, causing a 5 µm empty band in top of the images.
    Both zone plates have a 100 nm outer zone. 
    A: Wavefield at 7.6 keV produced by a zone plate with a 200 µm diameter. 
    The first order focus lies around 123 mm.
    B: Magnification of A around the first order focus.
    C: Wavefield at 7.6 keV produced by a zone plate with a 20 µm diameter. 
    The first order focus lies around 12.3 mm.
    C covers an area in space of the same size as B for comparison.
    D: Wavefields coming from the 20 µm zone plate at 7.6 keV, 8.1 keV and 8.6 keV, respectively combined into a single RGB image, showing the different focal points for different energies.
    }
    \label{fig:radial_zp_simulation}
\end{figure*}

Although $\epsilon$ does not depend on $\Delta_\mathrm{FZP}$, this parameter is still constrained by the setup since it determines the numerical aperture, which, for any given angular sampling requirement, is related to the width of the detector in order to preserve optimal beam spreading.
Therefore, to reduce $D_\mathrm{FZP}$ we have to reduce the number of zones $N_\mathrm{FZP}$, a number that scales to good approximation with $D_\mathrm{FZP}$ when $\Delta_\mathrm{FZP}$ is kept fixed. 
For example, for $\Delta_\mathrm{FZP}=100$ nm this means going from 500 zones in a 200 µm diameter FZP to 50 zones in a 20 µm diameter FZP. 
Despite the substantial reduction, computational wave propagation \cite{goodmanIntroductionFourierOptics2017}, shown in figures \ref{fig:radial_zp_simulation}A-C, suggests that the wavefield behind a smaller FZP looks sufficiently similar to that of a larger one in terms of points \ref{point:probe_control} (it is bounded in space with a sharp intensity drop-off), \ref{point:beam_spreading} and \ref{point:high_spatial_frequencies} (it converges and diverges) discussed in section \ref{sec:fzp_mono_ptycho}.
Furthermore, the diffraction efficiency, i.e. the fraction of the incoming intensity that ends up in each diffraction order, does not depend on the number of zones \cite{paganinCoherentXRayOptics2006}.
Figure \ref{fig:radial_zp_simulation}D serves to illustrate the energy dependent focal length of equation \ref{eq:focal_length}, using different colors to represent different energies in a single image.
Note that the low intensity center region in figures \ref{fig:radial_zp_simulation}B and \ref{fig:radial_zp_simulation}D, as caused by the comparatively large central stop, results in a ring-shaped rather than a circular probe.
To maintain the necessary redundancy between neighboring scan steps, the illuminated parts of the probe must overlap with itself, which means the step size must be small enough to accommodate this ring overlap.

As mentioned in section \ref{subsec:achromatic}, the chromaticity of the FZP, in combination with an OSA, can also be used to filter out the uninteresting parts of the spectrum to make more efficient use of the detector.
A sequence of small FZPs with different diameters then, can provide tunable spectral filtering through chromaticity, depending on the width of the spectrum that one would like to capture on the detector.
An important downside of a reduced FZP diameter is that the smaller surface area captures and focuses fewer photons.
However, when using hyperspectral detectors with a very low count rate capability, this is not a limiting factor.
Another point of attention is the increased stability demanded by smaller optics.
The small FZP strategy has recently been used by our group in a broadband ptychography experiment \cite{stolpBroadbandPtychotomographyHyperspectral2025}.

\section{Conclusion}
This article has explored several pre-sample optical designs and assessed their suitability for extending the permissible bandwidth of the broadband x-ray ptychography approach. 
The results demonstrate an ability to both increase the collection efficiency at synchrotron sources and to begin performing these experiments with the latest laboratory x-ray sources. 
The immediate impact is greatest in the laboratory environment --- where we expect to be able to start extracting high-resolution images with elemental sensitivity.  

Among the investigated strategies, the FZP with a reduced diameter appears to be the most effective, increasing both the bandwidth compared to a larger zone plate, and the collection efficiency compared to a pinhole. 
This strategy has been confirmed to work in a recent broadband ptychography experiment with a synchrotron source \cite{stolpBroadbandPtychotomographyHyperspectral2025}, where the increased efficiency allowed a first demonstration of hyperspectral ptychographic tomography and the increased bandwidth allowed to cover both the Nickel and Copper absorption edges.  

It should be noted that matching the diameter of the zone plate with the coherence length of the illumination is an important factor in x-ray ptychography, especially at low brilliance sources, as a mismatch can result in a large drop in the collected coherent flux. 
While a reduced FZP diameter may not always be ideal at high brilliance sources, we remain a long way from being limited by the level of coherent flux in that environment. 
The extended sources of laboratory sources and lower brilliance beamlines are a better match, and here a reduced FZP diameter keeps the experimental setup compact and portable.

However, in practice an efficiency loss emerges due to the central stop.
The size of the central stop is connected to the size of the OSA and both these parameters are limited by their production methods.
Furthermore, reducing these parameters makes alignment increasingly difficult.
Therefore, reducing the FZP diameter results in an increasing part of its surface area being blocked by the central stop, reducing efficiency through that mechanism.
To increase the total coherent flux and reduce acquisition times, one could blaze a FZP \cite{yunDevelopmentZonePlates1999}, although this increases the minimum possible outer zone width.
Another way to increase the throughput of the experiment is to use multi-beam ptychography \cite{yaoMultibeamXrayPtychography2020}.

Beyond physically increasing the flux, its use could be optimized by mitigating the "virtual" trade-off between spatial and energy resolution in post-processing. This could be achieved through spectral ptychography algorithms that exploit correlations across energy bins \cite{gaoSparseInitioXray2021a} or incorporate prior knowledge of the refractive index \cite{hiroseUseKramersKronig2017a}.  

Even for samples that are not inherently spectrally interesting, such algorithms make spectral data collection a valuable tool for maximizing source efficiency and extracting additional information within the same acquisition time.

Beyond physically increasing the flux, one could make better use of the existing flux through algorithms specialized for spectral ptychography, mitigating the current ``virtual'' trade-off between spatial and spectral resolution.
For example, one could exploit the strong correlations present across energy bins in spectral reconstructions, similar to what is done in e.g. \cite{gaoSparseInitioXray2021a} for a tomography reconstruction.
Another approach is to incorporate prior knowledge of the refractive index, either through specific information about the elements in the sample or through more general constraints such as the Kramers-Kronig relations \cite{hiroseUseKramersKronig2017a}.  
With such spectral algorithms, even when analyzing a sample that is not inherently spectrally interesting, collecting spectral data maximizes the source efficiency.

\begin{acknowledgments}
The authors acknowledge Fonds Wetenschappelijk Onderzoek (FWO-Flanders) for Research Project 3G010820 and SBO Project 3179I12018, Interreg Vlaanderen Nederland for Smart*Light 2.0 project and Ghent University Special Research Fund (BOF) for grant BOF/STA/202202/011. 
Diamond Light Source is acknowledged for the funding and beam time provided under proposals 
MG29218
, MG23409
, MG32637 
and MG34164. 
Frederic van Assche is acknowledged for developing the SpeXIDAQ software \cite{vanasscheSpectralXrayImaging2021} used for the creation and manipulation of spectral datasets.
\end{acknowledgments}

\section{Supplementary Material}
The supplementary material contains a larger version of figure \ref{fig:attenuator_correction_comparison} that includes the object modulus reconstructions, together with their own line profile.

\section*{Data availability statement}
The data that support the findings of this study are available from the corresponding author upon reasonable request.

%

\end{document}